# Generalized Langevin Mode Analysis (GLMA) for Local Density Fluctuation of Water in an Inhomogeneous Field of a Biomolecule

Fumio Hirata, Institute for Molecular Science*

*Professor Emeritus

**Abstract**

An idea to analyze the fluctuational mode or/and spectrum of water molecules in an *inhomogeneous* environment around a biomolecule such as protein and DNA is proposed based on the generalized Langevin equation (GLE) for water interacting with a biomolecule. The idea is to solve an eigen-value problem of the collective frequency matrix of water, which has a physical meaning similar to the Hessian matrix in the case of a connected harmonic-oscillator. The diagonalization of the matrix gives the mode and frequency of the density fluctuation of water around the biomolecule. The treatment is a natural extension of that made for water in a *homogeneous* environment by Chong and Hirata in 2008, in which the authors have found the four modes for density fluctuation; one for an *acoustic* mode, and the other three for *optical* modes. It is the point of the new approach how the mode and frequency of the density fluctuation are modulated by the field from the biomolecule. The method provides a way to distinguish the mode of fluctuation with *color*.

A few possible applications of the theoretical approach to problems of the life science including medicine are also proposed.

**I. Introduction**

Density fluctuation and dynamics of aqueous solution in an inhomogeneous environment around a solute molecule has been of great interest in the physical chemistry.

The topics attracted scientists in earlier stage of the solution chemistry, in which the ion hydration around alkali-halide ions are of great interest. [1,2] Many physicochemical measures have suggested that the fluctuations of water molecules in the hydration shell around $Li^+$ and $K^+$ are qualitatively different: for examples, water molecules in the first hydration shell around $Li^+$ is less mobile compared to those in bulk, while those around $K^+$ is relaxing even faster than bulk water. Such phenomena were characterized by the terminologies as "structure maker and breaker" in the seminal papers by Frank and Wen, and the "positive and negative hydration" by Samoilov. [1,2] Those findings have unambiguously indicated that the fluctuation of water around an ion depends on the inhomogeneous environment around an ion, in this case, the electrostatic interaction that depends on the distance from the central ion.

Theoretical characterization of the phenomena has been made by many authors based on the molecular dynamics (MD) simulation, and on the statistical mechanics of molecular liquids, or the XRISM theory. [3-5] Based on the XRISM theory, Chong and Hirata have made an analysis of the free energy profile or the potential of mean force (PMF) along the ion-water distance, and compared it with the similar profile corresponding to the water-water interaction: the height of the first maximum of PMF is defined as an *activation barrier*, $F_{ion}^{\dagger}$, of the diffusive motion of a water molecule around an ion. [5,6] The authors have defined a quantity, $\Delta F_{ion}^{\dagger} \equiv F_{ion}^{\dagger} - F_{water}^{\dagger}$, as a measure to discriminate whether the hydration around the ion is "positive" or "negative", where $F_{water}^{\dagger}$ is the activation barrier in bulk water. The theoretical results were in good harmony with the experimental characterization of ion hydration, "structure maker or breaker", or "positive or negative hydration", made by the earlier authors. [1,2] Namely, those ions that are classified as the "positive hydration" have exhibited $\Delta F_{ion}^{\dagger} > 0$, while the ions categorized as "negative hydration" have shown $\Delta F_{ion}^{\dagger} < 0$. [5] It is the simplest case in which inhomogeneity of the potential field around a solute molecule makes the density fluctuation and dynamics of water qualitatively different. The analysis makes an essential point to such phenomena in which the fluctuational mode of a water molecule is highly dependent on the inhomogeneous field where the molecule is placed.

Recently, an interesting finding was made by Suzuki and his coworkers with respect to the density fluctuation of water molecules around a protein based on the dielectric-relaxation spectroscopy. According to their measurement, the relaxation rate of water is faster around some amino-acid residues compared to those around other residues, and the faster water-molecules are playing important roles in the activity of the protein, in that case, to originate a driving force of the sliding motion of myosin on the actin filament. The faster water-molecules are referred to as "*hypermobile water*" by the authors. The study provides a theoretical motivation to analyze the local density-fluctuation and dynamics of water molecules around amino acids of protein. [7,8]

Theoretical characterization of the *fluctuational mode* of water in homogeneous environment has been made by Chong and Hirata while ago, combining the generalized Langevin equation (GLE) for the density fluctuation with the XRISM theory. [9-13] Along the course in the theoretical development, there was a conceptual shift concerning the physics of dynamics, which is worthwhile to be noted here. The temporal fluctuation or dynamics of water molecules described by the XRISM/GLE theory is primarily concerned with interaction-sites or atoms (O,H) of a molecule. [10] According to such a physical picture, the dynamics of water is viewed as a correlated *translational*-motion of atoms (O,H,H) in a molecule. It is quite different from a more conventional picture for the dynamics of water, casually called the "rot-translational" model in which the rotational and translational motion of a molecule are described separately with the cartesian and the angular coordinates. [13,14] In order to translate the interaction-site picture into the rot-translational one, they have made an eigen-value analysis of the "collective frequency matrix" appearing in the XRIM/GLE, that characterizes the *mode* of the density fluctuation of water. One of the four characteristic-frequencies were assigned to the translational motion of water molecules, referred to as "acoustic mode". The remaining three modes were assigned to the rotational motion around the three principal axes, which are called "optical mode". The sound velocity estimated from the dispersion relation concerning the translational motion, or the plot of the frequency ($\omega$) against the wave number ($k$) at $k$=0, was ~1500 m/sec, that is roughly the sound velocity of water at the room temperature. [9,10]

The XRISM/GLE theory was applied successfully to several irreversible phenomena of molecular liquid, such as the dielectric relaxation, viscosity, and sound absorption, combining with the *linear response theory* to take account for the perturbation applied to the system, such as an electric field and a sound pressure. [15-18]

In the present study, we propose a simple idea to extend the analysis made for pure water in a *homogeneous* environment to that in an *inhomogeneous* environment such as surface of protein. For that purpose, we employ the generalized Langevin theory combined with the RISM/3D-RISM equations, proposed by Kim and Hirata in 2012. [19] The authors have derived essentially two generalized Langevin equations (GLE), one for the structural fluctuation of protein, and the other for the density fluctuation of water, which are mutually *conjugated* with each other. In order to find the frictional force acting on each atom of protein, one has to solve the GLE for water. On the other hand, he needs to solve the GLE for protein to calculate the density fluctuation of water around protein, since the density fluctuation depends on the temporal structure of protein. The equation for the structural fluctuation of protein has been employed for theoretical clarification of a variety of life phenomena including the protein folding. [20-23]

In the present paper, we use the GLE of water to analyze the frequency spectrum at a temporal structure of protein.

**II. Brief review of the GLE theory for a biomolecule in water.**

***Generalized Langevin Theory:*** The generalized Langevin theory for a biomolecule in water developed by Kim and Hirata is briefly reviewed for completeness of the description. [12,13,19]

The generalized Langevin equations (GLE) describes the time evolution of dynamic variables $\mathbf{A}(t)$ in the phase space, which is governed by the Liouville operator $iL$.

$$\frac{d\mathbf{A}(t)}{dt} = iL\mathbf{A}(t) \qquad (1)$$

The dynamic variables represent a set of few mechanical variables which are essential to describe the physics of interest.

The formal solution of Eq. (1) is given by,
$$\mathbf{A}(t) = \exp(iLt)\mathbf{A}(0) = \exp(iLt)\mathbf{A} \qquad (2)$$

It is the essential idea of GLE to project all the variables in the phase space onto the dynamic variables using a projection operator defined by,

$$P = \frac{\mathbf{a} \cdot (\mathbf{a},\mathbf{b})}{(\mathbf{a},\mathbf{a})} \qquad (3)$$

In the definition, (**a**,**b**) denotes the inner product of the vector **a** and **b** in the Hilbert space, which is defined in terms of canonical ensemble average of **a**\***b** as

$$(\mathbf{a},\mathbf{b}) = (\mathbf{a}*\mathbf{b}) = Z^{-1}\int d\Gamma\, \mathbf{a}*\mathbf{b}\exp(-H(\Gamma)/k_B T) \qquad (4)$$

where $H(\Gamma)$ denotes the Hamiltonian of the system of interest.

The generalized Langevin equation for the dynamic variables $A$ can be derived by projecting all other variables in the phase space onto $A$ using the projection operator, that reads

$$\frac{d\mathbf{A}(t)}{dt} = i\Omega \mathbf{A}(t) + \int_0^t \mathbf{K}(t-\tau)d\tau + \mathbf{f}(t) \qquad (5)$$

where $i\Omega$, $\mathbf{K}(t-\tau)$, and $\mathbf{f}(t)$ are defined as

$$i\Omega \equiv \frac{(\mathbf{A},\mathbf{A})}{(\dot{\mathbf{A}},\mathbf{A})}, \quad \mathbf{f}(t) \equiv \exp[i(1-P)]L(1-P)\dot{\mathbf{A}}, \quad K(t) = \frac{(\mathbf{f}(t),\mathbf{f}(0))}{(\mathbf{A},\mathbf{A})} \qquad (6)$$

Eq. (5) has the form of the Langevin equation, and $i\Omega$, **f**(*t*), and *K*(*t*) are the collective frequency matrix, the random force, and the friction kernel.

***Generalized Langevin equation for a biomolecule in water***: We apply the general theory to a biomolecule-water system at infinite dilution, the Hamiltonian of which is defined as,

$$H = H_0 + H_1 + H_2 \qquad (7)$$

$$H_0 = \sum_{i=1}^{N}\sum_{a=1}^{n}\left[\frac{\mathbf{p}_i^a \cdot \mathbf{p}_i^a}{2m_a} + \sum_{j\neq i}\sum_{b\neq a} U_0\left(\left|\mathbf{r}_i^a - \mathbf{r}_j^b\right|\right)\right] \quad \text{(solvent)}, \qquad (8)$$

$$H_1 = \sum_{\alpha=1}^{N_u}\left[\frac{\mathbf{p}_\alpha \cdot \mathbf{p}_\alpha}{2M_\alpha} + \sum_{\beta\neq\alpha} U_1\left(\left|\mathbf{R}_\alpha - \mathbf{R}_\beta\right|\right)\right] \quad \text{(solute)}, \qquad (9)$$

$$H_2 = \sum_{\alpha=1}^{N_u}\sum_{i=1}^{N}\sum_{a=1}^{n} U_{\text{int}}\left(\left|\mathbf{R}_\alpha - \mathbf{r}_i^a\right|\right) \quad \text{(solute-solvent)}, \qquad (10)$$

where $H_0$, $H_1$ and $H_2$ are the Hamilitonian of solvent, solute, and the the interaction between them, respectively. The subscript *i* and *j* specify molecules in solvent, *a* and *b* distinguish atoms in a water molecule, and Greek characters $\alpha$ and $\beta$ label atoms in the biomolecule. [19]

The following vector in the phase space was chosen for the dynamical variables.

$$\mathbf{A}(t) = \begin{pmatrix} \Delta \mathbf{R}_\alpha(t) \\ \mathbf{P}_\alpha(t) \\ \delta\rho_\mathbf{k}^a(t) \\ \mathbf{J}_\mathbf{k}^a(t) \end{pmatrix}, \qquad (11)$$

where $\Delta \mathbf{R}_\alpha(t)$ and $\mathbf{P}_\alpha(t)$ are the displacement or fluctuation of an atom $\alpha$ from its equilibrium position in the biomolecule and its conjugated momentum, defined respectively by

$$\Delta \mathbf{R}_\alpha(t) \equiv \mathbf{R}_\alpha(t) - \langle \mathbf{R}_\alpha \rangle \quad \text{and} \quad \mathbf{P}_\alpha(t) \equiv M_\alpha \frac{d\Delta \mathbf{R}_\alpha}{dt}. \qquad (12)$$

The $\delta\rho_\mathbf{k}^a(t)$ and $\mathbf{J}_\mathbf{k}^a(t)$ are the Fourier transform of the density fluctuation of atom $a$ of water molecules, defined by,

$$\delta\rho^a(\mathbf{r},t) = \sum_i \delta(\mathbf{r} - \mathbf{r}_i^a(t)) - \langle \rho^a \rangle, \qquad (13)$$

and its conjugated momentum, or the Fourier component of the current of atom $a$ in the solvent, defined by

$$\mathbf{J}_a(\mathbf{r},t) \equiv \sum_i \mathbf{p}_i^a \delta(\mathbf{r} - \mathbf{r}_i^a). \qquad (14)$$

Applying the projection operator defined by Eq. (3) to the Liouville equation, one finds essentially two GLEs, one for the biomolecule, the other for water.
Biomolecule:

$$M_\alpha \frac{d^2 \Delta \mathbf{R}_\alpha(t)}{dt^2} = -k_B T \sum_\beta (\mathbf{L}^{-1})_{\alpha\beta} \cdot \Delta \mathbf{R}_\beta(t) - \int_0^t ds \sum \Gamma_{\alpha\beta}(t-s) \cdot \frac{d\Delta \mathbf{R}_\alpha(s)}{ds} + W_\alpha(t) \qquad (15)$$

Water:

$$\frac{\partial^2 \delta\rho_a(\mathbf{k},t)}{\partial t^2} = -k^2 \sum_{b,c} J_{ac}(\mathbf{k}) \chi_{cb}^{(2)}(\mathbf{k})^{-1} \delta\rho_b(\mathbf{k},t) + \left( \frac{1}{N} \sum_{b,c} J_{ab}(\mathbf{k}) \int_0^t ds M_{bc}(\mathbf{k};t-s) \cdot \frac{d\delta\rho_c(\mathbf{k},s)}{dt} ds \right) - k^2 \Xi_a(\mathbf{k},t)$$

(16)

Eq. (15) describes the structural fluctuation of a biomolecule on the free energy surface including the solvation free energy. The first term in the right-hand side is concerned with the restoring force which is proportional to the displacement of atoms from an equilibrium structure. The second and third terms are the frictional and random forces, which are related each other by the fluctuation-dissipation theorem. All the terms have general forms described in Eq. (6), but atomistic details of the equation are omitted here, because we are just interested in the density fluctuation of water around an equilibrium conformation of a biomolecule.

Eq. (16) describes the density fluctuation of atoms of water molecules around a biomolecule. The first term in the right-hand side is interpreted as a *restoring force* which is proportional to the fluctuation of the atomic density from its equilibrium value. The second and third terms are the frictional and random forces, which are related each other by the fluctuation-dissipation theorem.

**III. Mode and frequency of the density fluctuation of water around a biomolecule.**

*Equation of motion:* It is of great interest to investigate how the fluctuational mode and frequency of water changes due to an inhomogeneous field of a biomolecule. The essential characteristic of the mode is embodied in the first term in the right-hand side of Eq. (16). The third term, or the random force, is related to the thermal excitation of the mode, while the second term, or the friction, governs the relaxation rate of the mode. Therefore, we just focus on the first term, and rewrite the equation in the real (**r**) space by,

$$\frac{\partial^2 \delta \rho_a(\mathbf{r},t)}{\partial t^2} = \sum_b \int \omega_{ab}^{(2)}(\mathbf{r},\mathbf{r}') \delta \rho_b(\mathbf{r}',t) d\mathbf{r}', \qquad (17)$$

where $\omega_{ab}^{(2)}(\mathbf{r},\mathbf{r}')$ is defined by

$$\omega_{ab}^{(2)}(\mathbf{r},\mathbf{r}') \equiv \frac{d^2}{d\mathbf{r}^2} \sum_c J_{ac}(\mathbf{r}) \chi_{cb}^{(2)}(\mathbf{r},\mathbf{r}')^{-1}. \qquad (18)$$

The equation can be interpreted in an analogy to a coupled *harmonic-oscillator* in which the restoring force is originated from the deviation of atoms from the equilibrium position, if one views the right-hand-side of Eq. (14) as a *force* restoring the average density of an atom "*a*" of a water molecule at the position, **r**, from its equilibrium state. And, $\omega_{ab}^{(2)}(\mathbf{r},\mathbf{r}')$ is regarded as the "force constant" of the *restoring force*.

*Principal-axis analysis:* In order to complete the analogy with the harmonic oscillator, it will be a rational strategy to perform the principal-axis analysis for the "force constant matrix" $\omega_{ab}^{(2)}(\mathbf{r},\mathbf{r}')$ in order to find the characteristic frequency as well as the mode of the density fluctuation.

Such an analysis for the density fluctuation in *pure water* has been already carried out by Chong and Hirata as was briefly mentioned in the introduction section. [9,10] By diagonalizing the matrix similar to $\omega_{ab}^{(2)}(\mathbf{r},\mathbf{r}')$, the authors have found essentially four characteristic modes, one for the translational oscillation, or the "acoustic mode", and the other three for the rotational vibration around

the principal axis, assigned to the "optical mode."

If one applies the similar approach to Eq. (15), or its Fourier transform $\omega_{ab}^{(2)}(\mathbf{k})$, we will find the eigen frequencies and vectors for water molecules as a function of wave number **k**. By taking the Fourier transform of the eigen frequency for the sound mode, for examples, $\omega_{sound}^{(2)}(\mathbf{k})$, to the **r**-space, one may find the 3D- distribution of the frequency, $\omega_{sound}^{(2)}(\mathbf{r})$, in an inhomogeneous environment around a biomolecule.

***Computational recipes:*** The frequency matrix to be diagonalized is written in Fourier k-space as,

$$\omega^{(2)}(\mathbf{k}) = k^2 \sum_{b,c} J_{ac}(\mathbf{k}) \chi_{cb}^{(2)}(\mathbf{k})^{-1} \qquad (19)$$

where $J_{ac}(\mathbf{k})$ is single molecular quantity, which is concerned with the kinetic part of the density fluctuation. The concrete form of the function has been already worked out by several authors, and the formula for water are reviewed in the appendix.[9,26-28] The function does not depend on the local field exerted by the biomolecule.

The density correlation matrix, $\chi_{ab}^{(2)}(\mathbf{k})$, represents the pair density correlation of two positions, the real space expression is defined as,

$$\chi_{ab}^{(2)}(\mathbf{r},\mathbf{r}') \equiv \left\langle \delta\rho_a^{(1)}(\mathbf{r}) \delta\rho_b^{(1)}(\mathbf{r}') \right\rangle. \qquad (20)$$

In the expression, $\delta\rho_a^{(1)}(\mathbf{r})$ and $\delta\rho_b^{(1)}(\mathbf{r}')$ are the density fluctuation of oxygen or hydrogen atom of a water molecule at the positions **r** and **r'** around the biomolecule. It should be noted that the density pair correlation function, $\chi_{ab}^{(2)}(\mathbf{r},\mathbf{r}')$, is essentially a four-body correlation function, since each atom *a* and *b* are subject to the force field from the biomolecule on top of the correlation between themselves.

We assume the following approximation to the density pair correlation,

$$\chi_{ab}^{(2)}(\mathbf{r},\mathbf{r}') \sim \left\langle \delta\rho_a^{(1)}(\mathbf{r}) \right\rangle \left\langle \delta\rho_b^{(1)}(\mathbf{r}') \right\rangle, \qquad (21)$$

where $\left\langle \delta\rho_a^{(1)}(\mathbf{r}) \right\rangle$ and $\left\langle \delta\rho_b^{(1)}(\mathbf{r}') \right\rangle$ are the density correlation function of H or O atoms of water molecules around the biomolecule, each of which can be readily calculated based on the RIMS/3D-RISM theory. It is also of interest to examine the following approximation,

$$\chi^{(2)}_{ab}(\mathbf{k}) \sim \chi^{(1)}_{a}(\mathbf{k})\chi^{(1)}_{b}(\mathbf{k}), \qquad (22)$$

where $\chi^{(1)}_i(\mathbf{k})$ (*i=a* or *b*) is the expression of the density correlation function $\left\langle \delta\rho^{(1)}_i(\mathbf{r}) \right\rangle$ in the Fourier k-space. The physics involved in the two approximation is not exactly the same. However, we believe the results are close enough, in particular, for water molecules which are subject to rather strong filed of a biomolecule such as hydrogen-bonds, because the corelation of those water molecules are primarily governed by the field from the biomolecule. In any way, those functions can be evaluated readily by means of the RISM/3D-RISM theory for a fixed conformation of a biomolecule. In fact, $\left\langle \delta\rho^{(1)}_i(\mathbf{r}) \right\rangle$ has been calculated for many biomolecular systems, including protein and DNA in aqueous solutions, and the results have been used for calculating binding free energy of a variety of chemical compounds, including those of drug, to a cavity of protein molecules. [29-33]

The spectrum primarily obtained from the theory is a three dimensional plot of the frequency at a position **r** around a biomolecule. So, the spectrum so obtained may not look llike an ordinary one measured by the experimental sepctroscopy, in which the spectral intensity is plotted against the frequency or wave number. Nevertheless, it can be translated into the usual spectra by calculating a hsitogram of the frequencies sampled from the region of interest, for example, inside a protein cavity.

**IV. Possible applications**

In the previous sections, the author presented a theoretical idea to calculate the 3D-distribution of the frenquency and mode of the density fluctuation of water molecules which are subject to the field of a biomolecule in aqueous solution. The method makes possible to distingush the fluctuational mode and spectrum of water molecules subject to the field of a biomolecule. In the following, few examples of possible application of the theory is presented.

The most conventional method of spectroscopy in chemistry, such as infra-red (IR), ultraviolet (UV), X-ray, to identify the structure and dynamics of molecules is to use the structural and energetical information of a molecule, such as the electronic structure, the bond-length, the bond-angles, and so forth. When the molecule is put in a different enviroment such as surface of a protein, the molecular structure and energetics suffer from the external pertubation. It is this small change of the molecule caused by the small perturbation that is detected by the spectroscopy as a shift of the peak position as well as the shape of the spectra. Such a method may be useful if the spectral shift due to the perturbation is significant. However, it is certainly not the case for water molecules around the surface of a protein. The intermolecular interaction between protein and water is certainly few magnitude small than the interaction among atoms in a water molcule. So, the small change of spectra concerning the molecular structure due to the perturbation from protein will be burried in that of bulk water just as a noise.

Then, what type of fluctuational or dynamical modes will be sensitive to the perturbation from a protein? The answer to the question is the *translational and rotational motion of a water molecule as a whole*. For examples, if a water molecules is confined in a cavity of protein, the translational motion will be largely ristricted unlike that in the bulk. If the water molecule makes a hydrogen-bond with one of amino-acid residues, for example, then the rotational motion will be also largely confined. So, the spectra assigned to the rotational-vibration should be shifted toward higher frequency. Therefore, the spectrum of the rot-translational mode may play a role of a *finger print, or a marker,* of such a water molecule confined in a cavity of protein.

The best example to which the method may be applied is the *hypermobile water* around the actin-myocin complex, found by Suzuki and his coworkers by means of dielectric relaxation spectroscopy. [7] If the *hypermobile water* truely exists, the theory should predict the spectral shift toward the lower frequency side around the residues where the *hypermobile water* was detected by the experiment. So, it will be a good challenge for the method presented here.

Many medical applications are also conceivable. One of those is the identification of the binding complex of a protein and a drug compound. In many cases, a drug compound is bound to an active site of a target protein to kill a bactria or a virus, which is vital to survive for the germ. When the active site is free from a drug compound, it is usually filled by one or few water molecules depending on the structure and the amino-acid composition of the cavity. Those water molecules should be disposed from the cavity upon binding of a drug compound. One of such phenomena was observed theoretically in the case of oceltamivir bound to an influenza virus. [29] Those water molecules bound at an active-site may have some characteritic frequncy, that can be used as a *maker*. So, if we are able to measure the spectrum of the water molecules, we should observe a dramatic change in the spectrum at the *marker* frequency.

Another medical application to be conceivable is the diagnosis of a disease such as cancer. In fact, the Magnetic Resonance Imaging (MRI), the most popular medical device to detect the cancer cell, is using the fluctuational mode of water molecules as a maker of the cancer cell: the diffusive motion of water molecules in the cancer cell is faster than that in healthy ones. [30,31] While the MRI can detect a cancer in the resolution of cell level, the new method may be able to detect a cancer in molecular level, or in the resolution of protein and DNA molecules within a cell, since it may probe the local density fluctuation of water right next to a biomolecule.

**Appendix**:

Frequency moment matrix, $J_{ab}(k)$, in Eq. is defiend as the current-current correlation function as,

$$J_{ab}(k) \equiv \frac{1}{N}\langle \mathbf{J}^a_{-\mathbf{k}} \cdot \mathbf{J}^b_{\mathbf{k}} \rangle$$

where $\mathbf{J}^a_{-\mathbf{k}}$ is the Fourier transform of the current density defined by Eq. (14). The current density can be split into two contributions from the translational and rotational motions.

$$J_{ab}(k) = J_{ab}^{trans}(k) + J_{ab}^{rot}(k)$$

The expression for the each contribution is shown in Ref. 8 as follows,

$$J_{ab}^{trans}(k) = \frac{k_B T}{M} j_0(kl_{ab})$$

$$j_0(kl_{ab}) = \frac{\sin(kl_{ab})}{kl_{ab}}$$

$$J_{OO}^{rot}(k)/k_B T = \frac{1}{3}z_O^2\left(\frac{1}{I_x}+\frac{1}{I_y}\right),$$

$$J_{HH}^{rot}(k)/k_B T = \frac{1}{3}\left\{y_H^2\left(\frac{1}{I_x}+\frac{1}{I_z}\right)+z_H^2\left(\frac{1}{I_x}+\frac{1}{I_y}\right)\right\},$$

$$J_{OH}^{rot}(k)/k_B T = \frac{1}{3}(j_0+j_2)z_O z_H\left(\frac{1}{I_x}+\frac{1}{I_y}\right) - j_2 z_O^2 \sin^2\theta \frac{1}{I_x},$$

$$J_{HH'}^{rot}(k)/k_B T = \frac{1}{3}(j_0+j_2)\left\{-y_H^2\left(\frac{1}{I_x}+\frac{1}{I_z}\right)+z_H^2\left(\frac{1}{I_x}+\frac{1}{I_y}\right)\right\} - j_2 z_H^2 \frac{1}{I_x}.$$

where $j_m \equiv j_m(kl_{ab})$, the $m$th order spherical Bessel function.